\documentclass[aps,twocolumn,nofootinbib,superscriptaddress,amsfonts,floatfix]{revtex4-1} % for review and submission
\usepackage{graphicx,amsmath,amssymb,amstext}
\usepackage{amssymb,amsbsy,amsfonts,amsthm,color,bm}
\usepackage{xcolor}
\usepackage{epsfig}
\usepackage{graphicx}
\usepackage{subfigure}
\usepackage{hyperref}
\usepackage{cancel}
\usepackage{balance}
\usepackage{booktabs}
\usepackage{diagbox,multirow}
\usepackage[normalem]{ulem}
\graphicspath{{Figures/}}
\usepackage{hyperref}

\definecolor{rp}{cmyk}{0.2, 1, 0.6, 0}
\definecolor{rp}{cmyk}{0.2, 1, 0.6, 0}
\definecolor{green2}{cmyk}{0.27, 0, 1, 0.52}

\usepackage{color}

\hypersetup{
    colorlinks=true,       % false: boxed links; true: colored links
    linkcolor=green2,          % color of internal links
    citecolor=green2,        % color of links to bibliography
    filecolor=magenta,      % color of file links
    urlcolor=green2           % color of external links
}

\renewcommand{\eqref}[1]{Eq.~(\ref{#1})}
\newcommand{\bref}[1]{(\ref{#1})}
\newcommand{\Beq}{\begin{eqnarray}}
\newcommand{\Eeq}{\end{eqnarray}}

\def\lsim{\mathrel {\vcenter {\baselineskip 0pt \kern 0pt \hbox{$<$} \kern 0pt \hbox{$\sim$} }}}

\begin{document}

\title{General Relativistic Polarized Proca Stars}

\author{Zipeng Wang}
\email{zwang264@jhu.edu}
\affiliation{Department of Physics and Astronomy, Johns Hopkins University, Baltimore, MD 21218, USA}

\author{Thomas Helfer}
\email{thomashelfer@live.de}
\affiliation{Institute for Advanced Computational Science, Stony Brook University, Stony Brook, NY 11794 USA}

\author{Mustafa A. Amin}
\email{mustafa.a.amin@rice.edu}
\affiliation{Department of Physics and Astronomy, Rice University, Houston, Texas 77005, USA}

\begin{abstract}
Massive vector fields can form spatially localized, non-relativistic, stationary field configurations supported by gravitational interactions. The ground state configurations (p-solitons/vector solitons/dark photon stars/polarized Proca stars) have a time-dependent vector field {pointing in the same spatial direction} throughout the configuration at any instant of time, can carry macroscopic amounts of spin angular momentum, and are spherically symmetric and monotonic in the energy density. 
In this paper, we include general relativistic effects, and numerically investigate the stability of compact polarized Proca stars (linear and circularly polarized) and compare them to hedgehog-like field configurations (with radially pointing field directions). Starting with approximate field profiles of such stars, we evolve the system numerically using  3+1 dimensional numerical simulations in general relativity. We find that these initial conditions lead to stable configurations. However, at sufficiently large initial compactness, they can collapse to black holes. We find that the initial compactness that leads to black hole formation is higher for circularly polarized stars (which carry macroscopic spin angular momentum), compared to linearly polarized ones, which in turn is higher than that for hedgehog configurations. 
\end{abstract}

%\pacs{}
\maketitle

%%%%%%%%%%%%%%%%%%%%%%%%%%%%%%%%%%%%%%%%%%%%%%%%%%%%%%%%%%%%%%%%%%
\section{Introduction} \label{sect:intro}
Massive vector fields (or dark photons) can constitute all or part of dark matter. If their mass is $\lesssim 10\,\rm {\rm eV}$, the occupation numbers in typical astrophysical/cosmological settings are large enough to admit a classical field description. Their early universe production (for example, \cite{Graham:2015rva,Agrawal:2018vin,Co:2018lka,Dror:2018pdh,Bastero-Gil:2018uel, Long:2019lwl,Adshead:2023qiw,Kitajima:2023pby}), astrophysical/cosmological phenomenology, as well as direct and indirect detection strategies are being extensively explored (see \cite{Antypas:2022asj} for a recent review).
Numerical simulations investigating the nonlinear (and non-relativistic) gravitational dynamics of such fields in an astrophysical setting have been initiated recently \cite{Amin:2022pzv,Gorghetto:2022sue,Jain:2023ojg, Chen:2023bqy}.

Similar to the case of scalar fields, in the non-relativistic limit, one expects massive vector fields to form spatially localized, non-relativistic, stationary field configurations (solitons or Boson stars) supported by gravitational interactions. At any instant of time, such polarized Proca stars (also referred to as $p$-solitons, vector solitons, dark photon stars) have a spatially constant  orientation  of the field polarization throughout the configuration \cite{Adshead:2021kvl,Jain:2021pnk}. Depending on the polarization, they can carry macroscopic amounts of spin angular momentum \cite{Jain:2021pnk,Amin:2022pzv}. They are spherically symmetric in energy density, but not in the field configuration (but are node-free). Non-relativistic (fractionally) polarized solitons have been shown to from generically from cosmological, as well as astrophysical initial conditions \cite{Gorghetto:2022sue,Amin:2022pzv,Jain:2023ojg, Chen:2023bqy}.   

{General relativistic effects become necessary to consider if the vector field configurations become sufficiently compact.\footnote{Current simulations show that solitons forming from generic initial conditions are indeed fractionally polarized (with macroscopic spin) \cite{Jain:2023ojg}, and are typically too diffuse to warrant studying relativistic corrections. However, as such solitons accrete fields from their surrounding, they can become increasingly more compact.} Such an analysis is relevant for understanding the detailed nature of the compact configurations, including their maximal compactness, intrinsic spin, stability and deformability. These properties can be critical when considering (mergers of) such compact objects as gravitational wave sources \cite{Sanchis-Gual:2022mkk,CalderonBustillo:2020fyi}.}
\footnote{The merger of scalar boson stars and their  gravitational wave emission has been studied extensively in the literature, see for example \cite{Palenzuela:2006wp,Palenzuela:2007dm, Choptuik:2009ww,Bezares:2017mzk,Palenzuela:2017kcg,Helfer:2018vtq,Widdicombe:2019woy,Sanchis-Gual:2020mzb,Diamond:2021dth,Evstafyeva:2022bpr,Siemonsen:2023hko,
Jaramillo:2022zwg,Bezares:2022obu}.} In this paper, we study such polarized Proca stars within full numerical relativity. 

{We note that polarized Proca stars that we focus on here are unlike hedgehog-like field configurations. Hedgehog-like configurations have spherically symmetric field configurations with spatially varying field polarization, and have a node in their field profiles at the origin.   Hedgehog configurations have been studied in detail including general relativistic corrections, assisted by the spherical symmetry of the energy density and their field configuration \cite{Brito:2015pxa,Sanchis-Gual:2017bhw}. However, such hedgehogs are higher energy states of the field for a given mass compared to the polarized Proca stars/p-solitons mentioned above \cite{Adshead:2021kvl,Jain:2021pnk}.  Unlike the polarized Proca stars, hedgegogs have been shown to form only under a special set of spherically symmetric (field) initial conditions and evolution \cite{DiGiovanni:2018bvo}.}

{As far as general relativistic polarized Proca stars are concerned, the $m=1$ case was studied in \cite{Brito:2015pxa,Sanchis-Gual:2022mkk,PhysRevLett.123.221101} (for the complex vector field).\footnote{We thank William East and Nils Siemonsen for guiding us to relevant existing work here.} This case is similar to circularly polarized solitons/Proc stars with macroscopic angular momentum in \cite{Jain:2021pnk}. The analysis in \cite{Jain:2021pnk} (which includes circular, linear and fractionally polarized cases) is a non-relativistic analysis, where the underlying relativistic vector field can be real valued. To the best of our knowledge, general relativistic linearly polarized Proca stars (with negligible angular momentum) have not been studied in the literature before.}

In this study, we explore the behavior of complex-valued polarized Proca stars as their compactness is increased, which allows us to estimate a rough lower bound of the maximum compactness these stars. We numerically investigate the stability of both the linear and the circular polarization states, and compare them to hedgehog-like field configurations when general relativistic effects are included.\footnote{{The maximal compactness for the hedgehog configurations, and the circularly polarized ones was already provided in \cite{Brito:2015pxa}. The maximal compactness for  linearly polarized one, however, has not been provided.}}

We work in units where $\hbar=c=G_N=1$. In the captions we use the Planck mass $M_{\rm pl}\equiv 1/\sqrt{G_N}$, and occasionally even $\hbar$, to make the units explicit for clarity. We will use Greek letters ($\mu$, $\nu,\hdots$) to represent four-dimensional indices, and Latin letters
($i$,$j,\hdots$) to represent three-dimensional spatial indices. We work with the $-+++$ convention for the metric.

The rest of the paper is organized as follows. In section~\ref{sec:setup}, we provide the underlying model for complex Proca fields in general relativity. In this section we also describe the numerical relativity framework, construction of initial data and provide some details of the numerical set-up. In section~\ref{sec:results}, we present and discuss the  results of our simulations. We summarize our results and provide a taste of their implications in section~\ref{sec:summary}. Convergence tests and level of constrain violations are discussed in an appendix.

\begin{figure*}[h!]
  \includegraphics[width=0.98\textwidth]{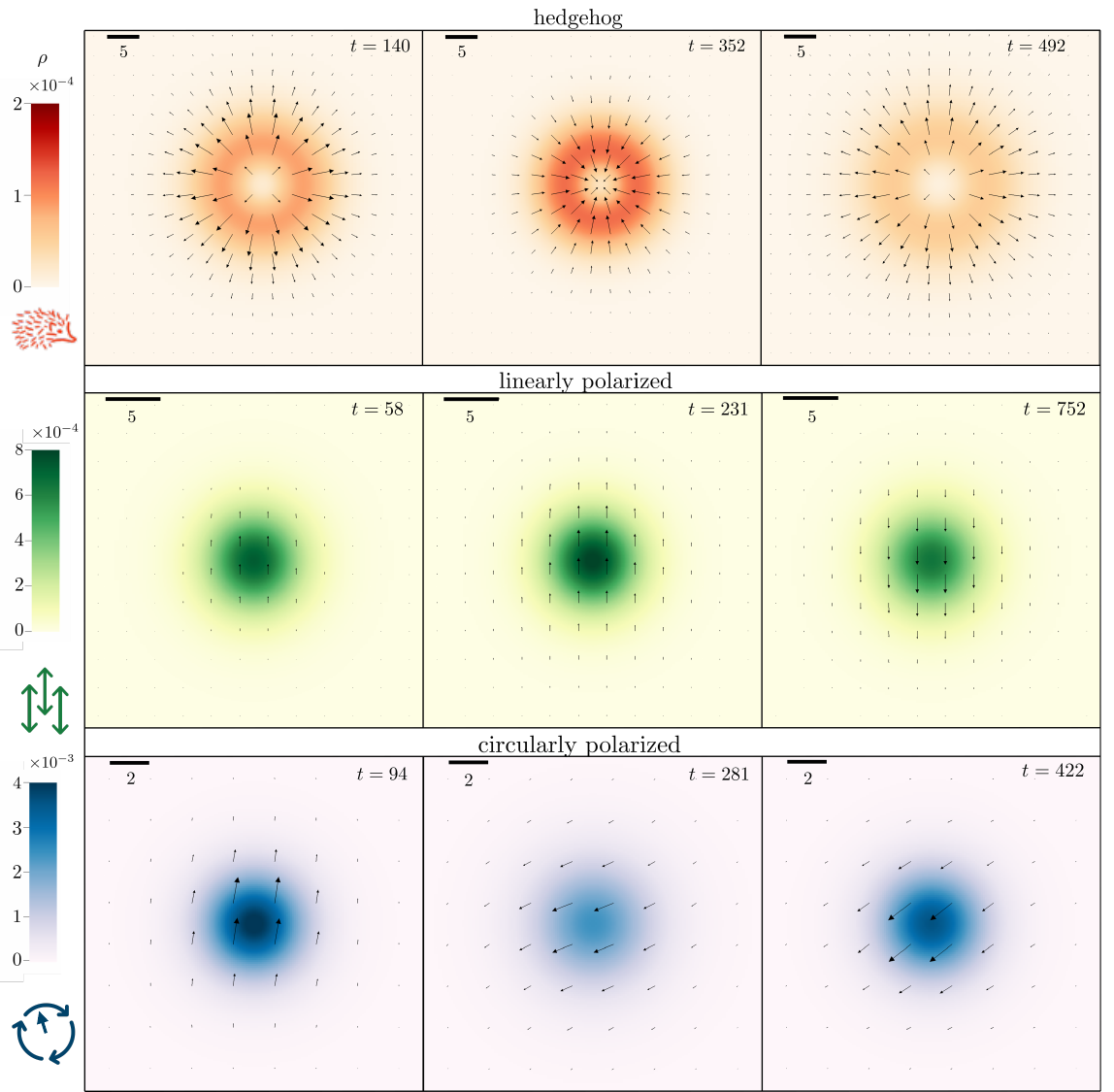}
  \caption{Simulations of the three types of Proca stars. Top row: a hedgehog configuration generated with  initial compactness $\mathcal{C}\approx 0.04$ ($\mu = 0.04m$) shown on the $x$-$y$ plane at three different times.  Middle row: a linearly polarized Proca star with $\mathcal{C}\approx 0.06$ ($\mu = 0.06m$) shown on the $x$-$z$ plane. The Proca field vectors are polarized in the $z$-direction. Bottom row: a circularly polarized star with initial compactness $\mathcal{C}\approx 0.08$ ($\mu = 0.10m$) shown on the $x$-$y$ plane. The energy density profiles, $\rho \times (M_{\rm pl}m)^{-2}$, are shown as color plots. The black arrows show the direction and relative magnitude of the real part of the spatial Proca vector field $\text{Re}(A_i)$. The black bars on top of each panel show the length scale of the plots in units of $m^{-1}$. The time shown is in units of $m^{-1}$. The (real part of) the vector field oscillates along the arrows in the top and middle panel, whereas it rotates in the bottom one (with period $T=2\pi m^{-1}$). Note that the time interval between snapshots is much longer than $T$; the changes in density profiles due to perturbations happen on these longer timescales.}
  \label{fig:screenshots}
\end{figure*}

\section{Setup and numerical methods}
\label{sec:setup}

\subsection{Proca field in general relativity}

We consider a  vector field $X^\alpha$ in general relativity, with an action \cite{Brito:2015pxa}
\begin{equation}
    \label{eqn:action}
    \mathcal{S} \!=\! \int  d^4x \sqrt{-g} \left( \frac{R}{16 \pi} - \frac{1}2 m^2 X_\alpha \bar{X}^{\alpha}
                    - \frac{1}4 F^{\mu \nu} \bar{F}_{\mu \nu}\right).
\end{equation}
Here, $R$ is the Ricci scalar, $g = \text{det}(g_{\mu\nu})$, $m$ is the mass of the vector field,
$\bar{X}^\alpha$ is the complex conjugate of the vector field ${X}^\alpha$, $F_{\mu \nu} = \partial_\mu X_\nu - \partial_\mu X_\nu$ is the field strength  tensor, and $\bar{F}_{\mu \nu}$ is its complex conjugate. 

Extremizing the action \eqref{eqn:action} with respect to variations in $g_{\mu \nu}$ yields the Einstein's field equation
\begin{equation}
    \label{eqn:EFE}
    G_{\mu \nu} = 8 \pi T_{\mu \nu}\,.
\end{equation}
Here, $G_{\mu \nu}$ is the Einstein tensor, and 
\begin{multline}
    \label{eqn:s-etensor}
    T_{\mu \nu} = \frac{1}2 \left( F_{\mu\rho} \bar{F}_\nu{}^\rho + \bar{F}_{\mu\rho} F_\nu{}^\rho \right)
                    -\frac{1}4F_{\rho \gamma} \bar{F}^{\rho \gamma}g_{\mu\nu}\\
                +\frac{m^2}2 \left( X_\mu \bar{X}_\nu + \bar{X}_\mu X_\nu 
                -g_{\mu \nu} \bar{X}^{\rho} X_{\rho}\right)\,
\end{multline}
is the stress-energy tensor associated with the Proca field. Similarly, extremizing the action \eqref{eqn:action} with respect to $X_{\mu}$ leads to the Proca field equation
\begin{equation}
    \label{eqn:4dproca}
    \nabla_{\mu} F^{\mu \nu} = m^2 X^{\nu}\,.
\end{equation}
The above equation with $m\ne 0$, along with the antisymmetry of $F^{\mu\nu}$, implies that the field $X^\nu$ must satisfy the Proca constraint equation
\begin{equation}
    \label{eqn:procaconstraint}
    \nabla_\nu X^\nu = 0\,.
\end{equation}
\eqref{eqn:EFE} and \eqref{eqn:4dproca} govern the evolution of the Proca field and spacetime. 

\subsection{3+1 decomposition}

Following \cite{Arnowitt:1959ah}, we foliate spacetime with spatial slices $\Sigma$ (with metric $\gamma_{ij}$), and connect the slices with each other with a lapse function $\alpha$ and shift vector field $\beta^i$. The spacetime metric can then be written as
\begin{equation}
    \label{eqn:ADM}
    ds^2 = -(\alpha^2-\beta^i\beta_i)dt^2 + 2 \beta_i dx^i dt + \gamma_{ij} dx^i dx^j\,,
\end{equation}
where $\beta_i\equiv \gamma_{ij}\beta^j$. 

The unit future-directed normal vector of $\Sigma$: $n_\mu = \left(-\alpha, 0, 0, 0 \right)$, and 
$P_{\mu}{}^{\nu} \equiv \delta_{\mu}{}^{\nu}+n_{\mu}n^{\nu}$ is the projection tensor that projects on to $\Sigma$ \cite{Croft:2022gks}. The extrinsice curvature of $\Sigma$ is $K_{ij} = -  \mathcal{L}_n \gamma_{ij}/2$ with $\mathcal{L}_{n}$ being the Lie-derivative along the normal vector $n_{\mu}$. A decomposition of the stress-energy tensor, \eqref{eqn:s-etensor}, adapted to this foliation of spacetime is
\begin{equation}\label{eqn:rhoSdecomp}
    \rho \equiv n_{\alpha} n_{\beta} T^{\alpha\beta},
    \quad\!\!\!\!
    S_i \equiv -\gamma_{i\alpha}n_{\beta} T^{\alpha\beta},
    \quad\!\!\!\!
    S_{ij} \equiv\! \gamma_{i\alpha}\gamma_{j\beta} T^{\alpha\beta}.
\end{equation}
The evolution equation for $\gamma_{ij}$, $K_{ij}$, $\alpha$, and $\beta^i$ can be obtained from the Einstein equation (\ref{eqn:EFE}); see \cite{Radia:2021smk} for their explicit form.

We also decompose the vector field as \cite{Zilhao:2015tya}:
\begin{equation}
    \label{eqn:ADMproca}
    X_\mu = A_\mu + n_\mu \varphi\,,
\end{equation}
where $\varphi=-n^\mu X_\mu$ is the component of the vector field normal to the spatial slice \cite{Zilhao:2015tya}, and $A_\mu=P_{\mu}{}^\nu X_\nu$ are the components along spatial slice.  The electric field associated with the Proca field is defined as:
\begin{equation}\label{eqn:Efield}
E_i\equiv P_{i}{}^{\mu}n^{\nu}F_{\mu\nu} \,.
\end{equation}
Under this decomposition, the Proca constraint, \eqref{eqn:procaconstraint}, becomes
\begin{equation}
    \label{eqn:procaconstraint3d}
    \mathcal{C_E} = D_i E^i - m^2 \varphi = 0\,.
\end{equation}
where $D_i$ is the covariant derivative corresponding to $\gamma_{ij}$.
The Proca evolution equation ,(\ref{eqn:4dproca}), yields \cite{Zilhao:2015tya}
\begin{align} \label{eqn:ADM_phi}
\partial_t \varphi &= - A^i D_i \alpha + \alpha (K\varphi - D_i A^i - Z) + 
    \mathcal{L}_{\beta} \varphi   \nonumber\,,\\
\partial_t A_i &= - \alpha (E_i + D_i \varphi) - \varphi D_i \alpha +
    \mathcal{L}_{\beta} A_i\, \nonumber, \\
\partial_t E^i &=  \alpha (K E^i + D^i Z + m^2 A^i + 
{D^k D^i A_k - D^k D_k A^i} ) \nonumber \\
%{D^k D_i A_k - D_i D^k A_k} ) \nonumber \\
                     &+ D^j \alpha (D^i A_j - D_j A^i)  +  \mathcal{L}_{\beta}
                     E^i\,, \\
\partial_t Z &= \alpha ( D_i E^i + m^2 \varphi - \kappa Z) + \mathcal{L}_{\beta} Z \nonumber\,.
\end{align}
In the above equation, $\mathcal{L}_\beta$ is the Lie derivative with respect to the shift vector, $K =\gamma^{ij}K_{ij}$ is the mean curvature, and $Z$ is an auxiliary variable introduced to keep $\mathcal{C_E}$ minimized during evolution \cite{Zilhao:2015tya,Hilditch:2013sba,Palenzuela:2009hx}.

\subsection{Initial data}\label{sec:initial data}
In this section, we describe the construction of the initial data, which approximates the full-GR solutions for three types of Proca stars.

\subsubsection{Proca field}
\label{sec:procafield}
As initial data, we use the field profiles for stationary Proca stars in the non-relativistic regime, $|\partial_i/m|\ll 1$. For details of this construction, see \cite{Jain:2021pnk}. The three types of Proca stars under consideration have a spatial vector field
\begin{equation}
\label{eq:InitialField}
    {\bm A}(\tilde{t},\tilde{\bm r})=e^{i\tilde{t}}\frac{\mu}{m}
    \begin{cases}
   f^{\rm lin}(\tilde{r}) \hat{\bm z} &\textrm{linearly polarized},\\
     f^{\rm cir}(\tilde{r})\dfrac{\hat{\bm x}+i\hat{\bm y}}{\sqrt{2}}&\textrm{circularly polarized},\\
    f^{\rm hh}(\tilde{r})\hat{\bm r} &\textrm{hedgehog}.
    \end{cases}
\end{equation}
Here, $\mu$ is the effective chemical potential with $\mu/m\sim |\partial_i^2/m^2|\ll 1$ in the non-relativistic limit, and the rescaled coordinates $\tilde r \equiv \sqrt{m\mu }r$, $\tilde{t}\equiv ( 1 - \mu/m)mt$. The profiles $f^{\rm lin}$, $f^{\rm cir}$ and $f^{\rm hh}$ are approximately given by 
\begin{align}
    f^{\rm lin}(\tilde{r})&=f^{\rm circ}(\tilde{r})\approx \frac{1.94}{(1+0.073 \tilde{r}^2)^4},\\
    f^{\rm hh}(\tilde{r})&\approx \frac{0.76\tilde{r}}{(1+0.0096 \tilde{r}^2)^{16}}.
\end{align}
These are fitting formulae for the profiles. More accurate profiles can be obtained by numerically solving the corresponding profile equations \cite{Jain:2021pnk}. Our fits deviate from these numerically obtained profiles by $\sim 5\%$.

To specify the electrical field \bref{eqn:Efield}
\begin{equation}
    \label{eqn:initialE}
    E_i = \gamma^{-\frac{1}{3}} \left( \partial_i \varphi- \partial_0 A_i \right)\,,
\end{equation}
where $\gamma \equiv \text{det}(\gamma_{ij})$. On the initial slice, we use \eqref{eq:InitialField} to obtain $\partial_0A_i$. We ignore the $\partial_i\varphi$ since it is suppressed by $|\partial_i|/m$. This $E^i=\gamma^{ij}E_j$ can then be used in \eqref{eqn:procaconstraint3d} to obtain $\varphi$.\footnote{We need not have ignored $\varphi$, and could have solved for it using \eqref{eqn:procaconstraint3d}, together with the Hamiltonian and momentum constraint equations. We found that this procedure was not numerically stable.}

\subsubsection{Spacetime}\label{sec:spacetime}
On the initial slice, the gauge functions are assumed to be trivial, $\alpha = 1$  and $\beta^i = 0$. The spatial metric is assumed have conformal flatness: $\gamma_{ij} = \psi^4 \delta_{ij}$, where $\psi = [\text{det}(\gamma_{ij})]
^{\frac{1}{12}}$ is the conformal factor of the metric.

The fields from the previous subsection and the metric,  must satisfy the Hamiltonian and momentum constraint 
\begin{align}
  &\mathcal{H} \equiv  R - K_{ij}K^{ij} + K^2 - 16 \pi \rho =0\,,\label{eqn:Hamcon}\\
&\mathcal{M}^i \equiv D_j\left( K^{ij} - \gamma^{ij} K \right) - 8 \pi S^i =0\,,\label{eqn:Momcon}
\end{align}
where $R$ is the three-dimensional Ricci scalar of $\gamma_{ij}$. 

To solve the constraint equations, we follow the conformal-transverse-traceless (CTT) formalism (see \cite{Baumgarte:2010ndz} chap.~3 and appendix B). We decompose the extrinsic curvature as $K_{ij} = \psi^{-2}\bar{A}_{ij} + \gamma_{ij}K/3$, where
$\bar{A}_{ij}$ is the trace-free part of the extrinsic curvature. We further assume zero mean curvature, $K=0$, on the initial slice. 
We decompose Eqns.~\bref{eqn:Hamcon} and \bref{eqn:Momcon} as 
\begin{align}
\label{eqn:CTTHam}
\Delta \psi + 
\frac{1}{8}\psi^{-7} \bar{A}_{ij}\bar{A}^{ij} = -2 \pi \psi^5 \rho\,,\\
\label{eqn:CTTMom}
(\Delta_L W)^i  = 8 \pi \psi^{10} S^i\,,
\end{align}
where $\Delta \psi = \partial^k \partial_k \psi$ is the flat Laplacian of the conformal factor, $W^i$ is the vector potential of $\bar{A}_{ij}$, and $(\Delta_L W)^i = \partial^j \partial_j W^i + \frac{1}{3} \partial^i\partial_j W^j$ is the flat vector Laplacian of $W^i$. 

We utilize a {\sc GRChombo}-based elliptical solver that has adaptive mesh refinement support 
to solve Eqns.~\bref{eqn:CTTHam} and \bref{eqn:CTTMom} with $\rho$ and $S_i$ given by \eqref{eqn:rhoSdecomp} based on the vector field profiles. Additionally, the solver requires an initial guess for the conformal factor $\psi$, which was set to be the conformal factor of the full-GR stationary hedgehog stars in \cite{Brito:2015pxa}. This solver improves this initial guess interatively, updating $\psi$ and $W^i$ each time to reduce the Hamiltonian and momentum constraints.
At each iterative step, we update the $E^i$ components of the field according to \eqref{eqn:initialE} (without $\partial_i\varphi$), and then update the $\varphi$ component of the Proca field according to \eqref{eqn:procaconstraint3d}, to ensure that the Proca constraint is still satisfied under the updated conformal factor $\psi$. Note that since the Proca field distribution is compact, we use the boundary conditions $\psi=1$ and $W^i=0$ and put the boundaries far from the Proca star.

For the densest simulated Proca star generated with $\mu=0.10m$, we solve the equations with side length $L = 300 m^{-1}$ with the number of points $N=96$ on the coarsest level. We add three additional refinement levels enclosing the star, with the finest resolution as $\Delta = 0.2 m^{-1}$. Under these conditions, the procedure detailed above provides good convergence rates, with $\mathcal{H}$ sufficiently small. In the Appendix, we demonstrate convergence for $\mathcal{H}$ with different resolutions as an example. The momentum constraint $\mathcal{M}^i$ and the Proca constraint $\mathcal{C_E}$ converge in a similar fashion.\footnote{However, we observed that, for denser initial vector profiles with parameter $\mu>0.10m$, this procedure fails to converge.}

For convenience, we call the Proca profiles detailed in \ref{sec:procafield}, along with the spacetime metric solved with \eqref{eqn:CTTHam} and \eqref{eqn:CTTMom}, ``constructed Proca stars".

A few comments are in order regarding our constructed Proca stars. We expect our constructed Proca stars to be non-stationary. This is because the non-relativistic profiles we use will deviate from the true relativistic solutions as the compactness is increased. This is currently unavoidable for us because unlike hedgehog-like Proca stars \cite{Brito:2015pxa}, constructing a stationary solution at high compactness for polarized stars is difficult due of the lack of spherical symmetry in the field configuration (and a likely deviation from spherical symmetry in the energy density).  Furthermore, in our procedure, we used $K=0$ and conformal flatness, ignored the transverse traceless part of the $K_{ij}$ and chose trivial functions for $\alpha$ and $\beta^i$ (chosen as a numerical convenience) which might make the initially constructed Proca stars deviate even more from their Newtonian counterparts at low compactness.  These shortcomings can be thought of as adding initial perturbations to the possible stationary solution for each of the three stars.

\subsection{Extraction of mass and angular momentum}
Following \cite{Clough:2021qlv, Croft:2022gks}, we define the conserved mass ($M$) and the $z$ component of the angular momentum ($J_3$) as 
\begin{align}\label{eqn:new_measure}
    Q=Q_0+\int_0^t \mathcal{S} dt\,,
\end{align}
where $Q=M,J_3$, and
\begin{align}
    Q_0&\equiv\int_{\Sigma} d^3 x  \sqrt{\gamma} \, n_\nu \zeta^\mu {T_\mu}{}^\nu,\\
     \mathcal{S} &\equiv \int_{\Sigma} d^3 x  \sqrt{\gamma} \, \alpha T^\mu{}_\nu \nabla_\mu \zeta^\nu\,.
\end{align}
The above quantities differ for $Q=M,J_3$ in the choice of $\zeta^\mu$:
\begin{equation}
    \zeta^\mu=
    \begin{cases}
     (1,0,0,0) & \textrm{for}\quad Q=M\,,\\
     (0,-y,x,0) & \textrm{for}\quad Q=J_3\,.
    \end{cases}
\end{equation}
The explicit expressions for $M_0$ and $(J_3)_0$ are given by
\begin{align}\label{eqn:mass0def,angmom0def}
    M_0 &= \int_{\Sigma} d^3 x  \sqrt{\gamma} \left( \alpha \rho - \beta_j S^j \right)\,\\
    (J_3)_0&=\int_{\Sigma} d^3 x  \sqrt{\gamma} \left( yS_x - xS_y \right)
\end{align}
where we $\rho,S_i$ are defined in \eqref{eqn:rhoSdecomp}.

\begin{figure}[t!]
  \includegraphics[width=0.47\textwidth]{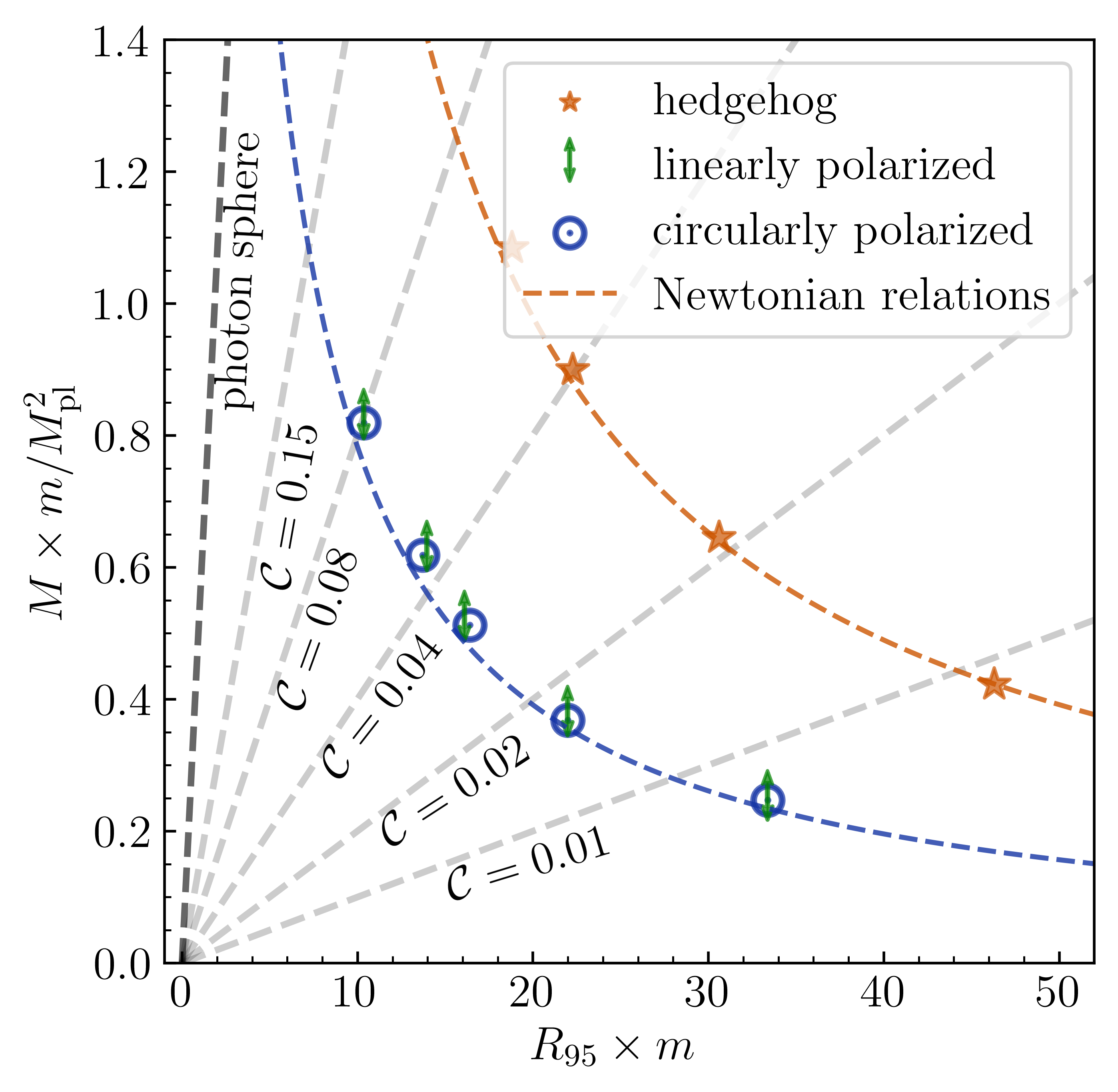}
  \caption{Initial mass and radius of Proca stars generated from non-relativistic profiles. The dots show the Proca stars with Proca field generated from Sec.~\ref{sec:procafield} and the spacetime metric solved from the CTT procedure.
  The orange and blue dashed lines show the mass-radius curve as predicted in the non-relativistic limit and under Newtonian gravity.}
  \label{fig:Newtonianfit}
\end{figure}

In Fig. \ref{fig:Newtonianfit} we summarize the mass-radius relationship of the constructed Proca stars solved using the CTT procedure in Sec.~\ref{sec:spacetime}, using $Q$ in \eqref{eqn:new_measure} as the measure of mass. We show that the initial data we obtain for all three kinds of stars agree approximately with the mass-radius curve in the non-relativistic limit \cite{Jain:2021pnk}. The compactness 
\Beq\mathcal{C}\equiv \frac{M}{R_{\rm 95}},
\Eeq
of these stars ranges from $\approx 0.01$ to $0.1$. Here, $R_{95}$  is defined as the radius containing $95\%$ of the mass.
We note that the above measure for ``mass", is really a measure of the total energy, including the rest mass. It will agree with the rest mass (defined in the Newtonian solutions) at low compactness, but can show deviations at larger ones.

%%%%%%%%%%%%%%%%%%%%%%%
\begin{figure}[h]
  \includegraphics[width=0.47\textwidth]{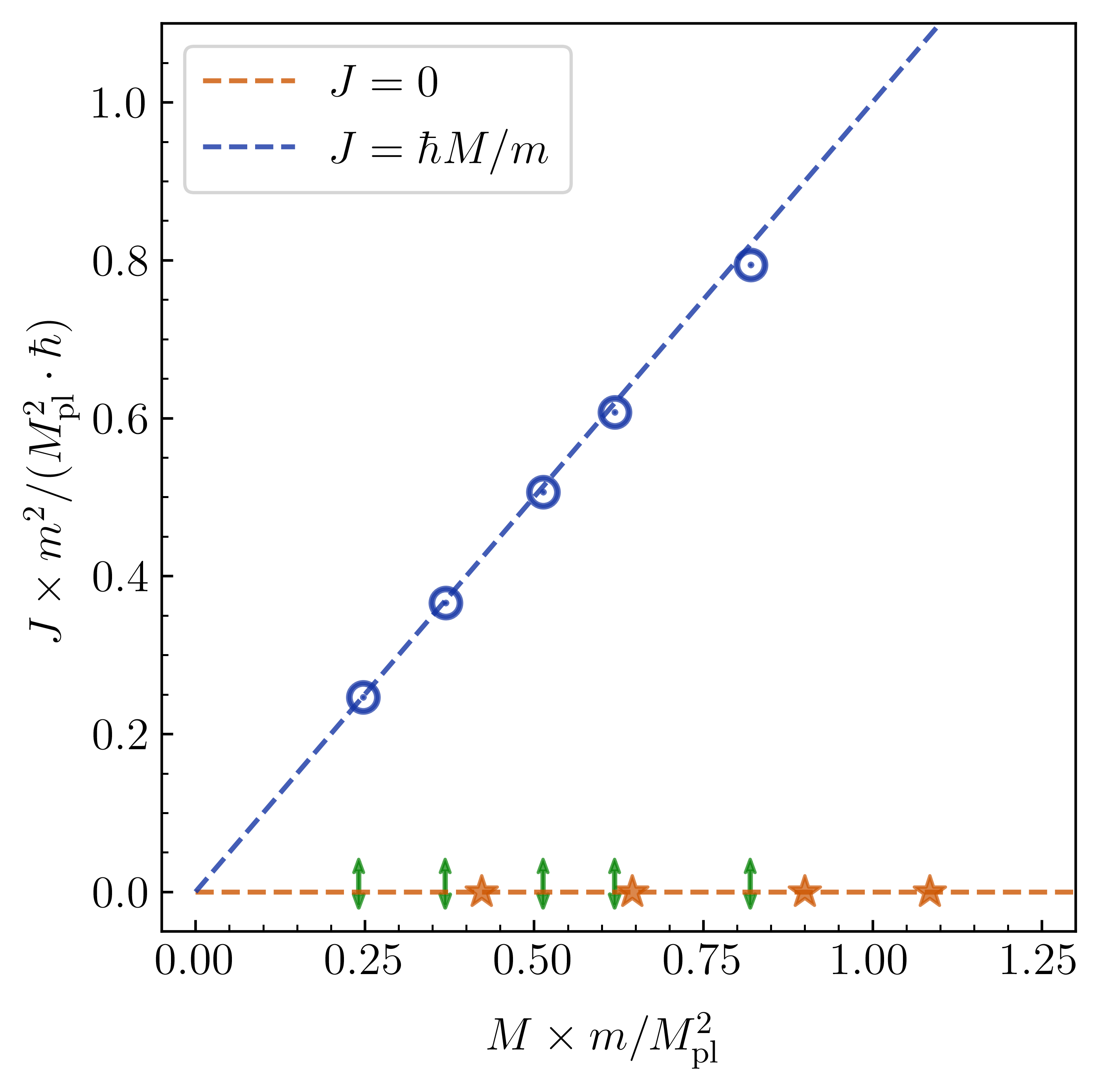}
  \caption{Total angular momentum against mass of the simulated Proca stars. For both the hedgehog stars and the linearly polarized stars, the extracted angular momentum from their respective simulations is consistent with zero. For the circularly polarized stars, the extracted angular momentum satisfies the relationship $J = \hbar M/m$, consistent with it being the spin angular momentum discussed in \cite{Jain:2021pnk}. 
  }
  \label{fig:JvsM}
\end{figure}

\section{Results \& Discussion}
\label{sec:results}
\begin{figure*}[ht]
  \includegraphics[width=1\textwidth]{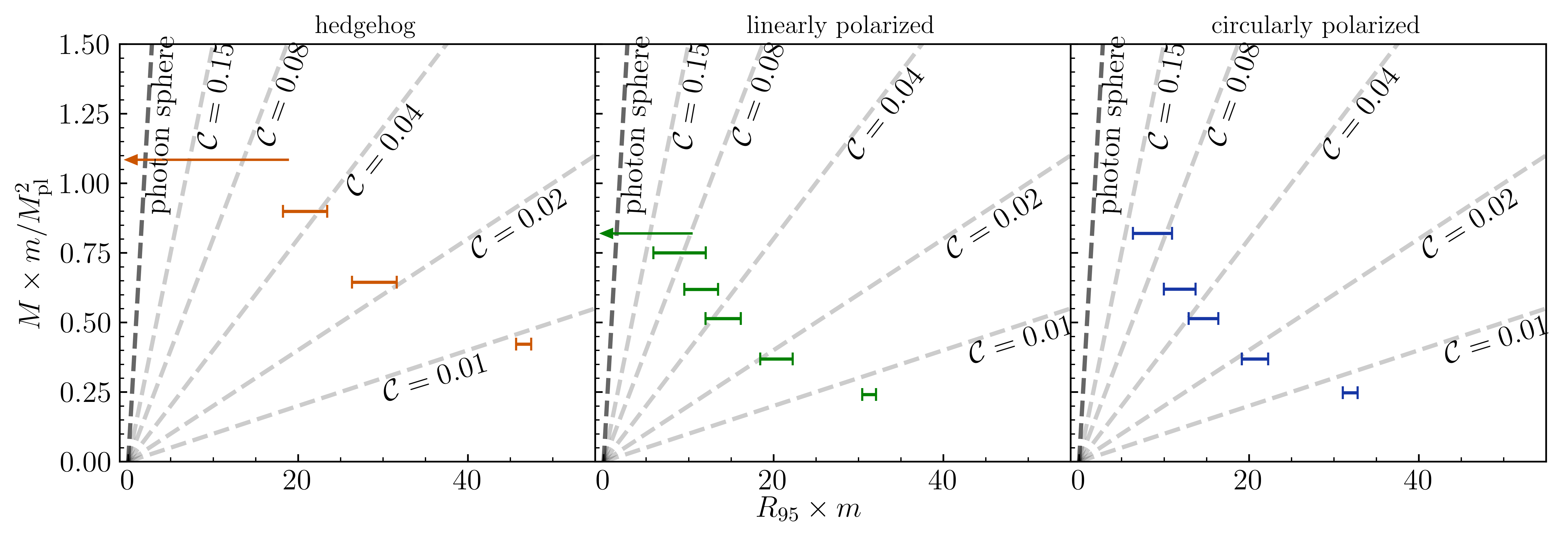}
  \caption{The mass-radius relationship of the simulated Proca stars. In all panels the bars show ranges of the radial changes observed in simulations (after an initial ``settling in" period of $400 m^{-1}$). For some stars, the bars are replaced by arrows, indicating that the Proca star collapses into a black hole. The grey dashed lines show the lines of constant compactness, and the dark grey line show the compactness of a black hole, with its photosphere as its radius, in isotropic coordinates. Left, Middle and Right panels show the results for four hedgehog Proca stars, six linearly polarized Proca stars, and  five circularly polarized stars respectively. For $\mathcal{C}\lesssim 0.1$, the middle and left panels demonstrate the stability of compact, gravitationally supported polarized stars. Near the upper bound of this range, hedgehogs collapse at the lowest initial compactness ($\mathcal{C}\approx 0.06$), followed by linearly polarized {($\mathcal{C}\approx 0.08$)}, and then (likely) circularly polarized stars {($\mathcal{C}>0.08$,} although we were unable to simulate collapse in circularly polarized stars). For non-collapsing polarized stars, the mean of the radial variations provides insight into the mass-radius relationship at these compactness. 
  }
  \label{fig:MvsR}
\end{figure*}
We simulated the evolution of  Proca stars (both polarized stars, as well as hedgehog-like configurations for comparison), using constraint-fulfilling initial Proca field profiles that are stationary under Newtonian gravity. The evolution times were approximately $140$ cycles  of the Proca field ($t_{\rm sim}\approx 900 m^{-1}$).

A sample evolution of the three different stars is shown in Fig.~\ref{fig:screenshots}. During their time evolution, all three stars exhibit radial oscillations in their density, but do not disperse away for the duration of the simulation. The period of this radial density oscillation is roughly $30 - 100$ times longer than period of vector field cycle ($T=2\pi m^{-1}$). The radial oscillations are likely excited due to the choice of initial data, see the last paragraph of section~\ref{sec:initial data}.

A summary of the  mass radius-relationship of the three types of Proca stars during the evolution (with varying initial compactess) is shown in Fig.~\ref{fig:MvsR}. For $\mathcal{C}\lesssim 0.1$, we see that all the stars show stable radial oscillations with the amplitude of the oscillations being smallest at lowest compactness. We take the survival of these stars for the duration of the simulation, with perturbations introduced by imperfect initial data, as  evidence of the existence of long-lived, compact, polarized Proca stars within full GR. 

In Fig.~\ref{fig:MvsR}, we can interpret the central values within the radial variations (horizontal ``error bars") for each $M$ as defining the  radius of a true stationary solution  at that mass. This should provide guidance in the construction of the stationary solution at large compactness for the linearly polarized stars, which is not known away from the Newtonian limit. However, we urge caution here, especially as compactness gets high. For the hedgehog and circularly polarized case, the mass radius relation is known at all compactness \cite{Brito:2015pxa}. 
{Using this known mass radius relationship for hedgehog configurations}, the radial variation does not include the expected radius for $\mathcal{C}\approx 0.04$. {We leave a more detailed comparison with known stationary mass-radius relationships, as well as the derivation of the stationary mass-radius curve for the linearly polarized case in the high compactness regime for future work.}

As the compactness approaches $\mathcal{C}=0.1$, we start seeing qualitatively different behavior for the three types of stars. We start seeing collapse to black holes in some types of stars. At $\mathcal{C}\approx 0.08$, the circularly polarized star shows large radial oscillations, but does not collapse to a black hole (left most point in the right panel of Fig.~\ref{fig:MvsR}. The linearly polarized one (middle panel), however does collapse at this same compactness. The hedghog star (left panel) collapses at an even smaller compactness of $\approx 0.06$ (where both linear and circularly polarized stars are stable). That is, the initial compactness range $(0,\mathcal{C}$) where  hedgehogs are stable is smaller compared with that of the polarized stars. Between the two polarized stars, the stable region of the linearly polarized stars is smaller than that of the circularly polarized ones. 
 We were unable simulate any type of Proca stars with  $\mathcal{C} > 0.08$ because the constraint equations {solver} Eqns.~\bref{eqn:CTTHam} and \bref{eqn:CTTMom} failed to converge {for these configurations}. 
\\ \\
\noindent{\bf{Limitations}}: We cannot control the magnitude of the perturbations around the stationary solution induced by the imperfect initial conditions. Therefore, it is possible that the perturbation is larger in the case of hedgehogs, which might be a confounding factor leading to collapse to black holes at smaller compactness. For this reason, we cannot ``prove" that that polarized stars are more stable than the hedgehog configuration. A controlled quantitative analysis will be possible, after the stationary solutions (like the $m=1$ case in \cite{Brito:2015pxa}) of polarized Proca stars are found at high compactness.

At large field amplitudes corresponding to highest  compactness explored here ($|A_i|\sim 0.1M_{\rm pl}$, the self-interactions of the vector field might not be ignorable. While polarized Proca stars with self-interactions for relatively small compactness have been explored in the literature \cite{Zhang:2021xxa,Jain:2022kwq,Jain:2022agt,Zhang:2023fhs}, the large amplitude here might bring additional complications, see \cite{Mou:2022hqb,Clough:2022ygm,Coates:2022qia,Coates:2022nif}.
%%%%%%%%%%%%%%%%%%%%%%%%%%%%%%

%%%%%%%%%%%%%%%%%%%%%%%%%%%%%%
\section{Summary \& Implications}\label{sec:summary}
%%%%%%%%%%%%%%%%%%%%%%%%%%%%%%
We simulated two types of polarized Proca stars (linear and circularly polarized), along with hedgehog-like Proca stars for comparison, using general relativistic field equations. The initial conditions were based on field profiles of related Proca star solutions in Newtonian gravity \cite{Jain:2021pnk}(see our Fig.~\ref{fig:Newtonianfit}), scaled to a higher compactness. 

Our key results are as follows (see Fig.~\ref{fig:MvsR}):
\begin{itemize}
\item We provided evidence that high-compactness polarized stars can be stable for $\mathcal{C}\lesssim 0.1$. 
\item As we increase the initial compactness from approximately $0.01$ to $0.1$, the linearly polarized, circularly polarized, and hedgehog stars evolve away from their initial configurations and towards new, and slightly different fixed points.
\item At sufficiently high compactness, some types of stars collapse to black holes. We found that circularly polarized stars avoid collapse to black holes up to higher initial compactness than linearly polarized ones, which in turn avoid collapse up to a higher initial compactness than hedgehog-like stars. {The large intrinsic spin angular momentum of circularly polarized stars (see Fig.~\ref{fig:JvsM}) might be playing a role in their relative robustness to collapse.}
\end{itemize}
For circularly polarized stars, we did not observe collapse to a black hole up to $\mathcal{C}=0.08$. We were unable to simulate stars with initial compactness $\gtrsim 0.08$ due to numerical limitations. An improved procedure for constructing the initial data which allows for control of perturbations away from the stationary solution is needed. This can be done using an improved initial data formulation such as the one in \cite{Brito:2015pxa}.

We hope our findings provide new phenomenology that can be incorporated in the search for ``exotic" compact objects \cite{Cardoso:2019rvt}  through gravitational and electromagnetic radiation. Polarized Proca stars can form in dark photon/ vector dark matter fields \cite{Gorghetto:2022sue,Amin:2022pzv,Jain:2023ojg, Chen:2023bqy}, potentially providing access the nature of the dark sector. 

For the purpose of gravitational wave physics, both the increased compactness, and the polarization of the stars, can have important implications. The increased maximal compactness of polarized stars in this paper (compared to, for example, hedgehog stars), suggests that they will get  closer before they merge, resulting in the emitted gravitational radiation being different from hedgehog stars. The polarization of the star can also impact the dynamics of the binary merger of such stars through finite size effects such as tidal deformability $\Lambda \propto \mathcal{C}^{-5}$ \cite{poisson_will_2014}, before and during merger.\footnote{See  \cite{Herdeiro:2020kba} for the tidal deformability of hedgehog Proca stars. More generally, the relevance of large tidal deformabilty of boson stars for gravitational wave emission is discussed in detail in \cite{Chia:2023tle} (also see \cite{Cardoso:2017cfl,Sennett:2017etc}).} In addition, circularly polarized stars  with maximal intrinsic spin can lead to spin-orbit and spin-spin effects before they merge. During the final phase of the merger, the generated gravitational waves can also be directly impacted by the polarization state of the star. Analysis of mergers of compact scalar boson with angular momentum leads to rich dynamics (see, for example \cite{Siemonsen:2023hko,Sanchis-Gual:2018oui}). A similar analysis is warranted for polarized Proca stars; for related recent work see \cite{Sanchis-Gual:2022mkk,CalderonBustillo:2020fyi}.

We have focused on stars constructed out of complex valued Proca fields for convenience. Similar constructions can be carried out for real valued fields (which might have a different lifetime). As in the case of axion stars \cite{Hertzberg:2018zte,Levkov:2020txo,Amin:2020vja,Amin:2021tnq,Sanchis-Gual:2022zsr,Chung-Jukko:2023cow}, such polarized Proca stars can also emit electromagnetic radiation, with the novelty that the properties of the radiation now depends on the polarization state of the Proca star \cite{Amin:2023imi}(for effects on gravitational radiation, see \cite{Nakayama:2023jhg}).  In particular, the polarization patterns in the outgoing radiation could provide a new handle on the nature of the underlying dark fields. It could be interesting to construct multimessenger signals (gravitational and electromagnetic waves) from merging polarized Proca stars.  

%%%%%%%%%%%%%%%%%%%%%%%%%%%%%%%%%%%%%%%%%%%%%%%%%%%%%%%%%%%%%%%%%%%%%%%%%%%%%%%%%%%%%%%
\acknowledgments 
The authors acknowledge the Texas Advanced Computing Center (TACC) at The University of Texas at Austin for providing {HPC, visualization, database, or grid} resources that have contributed to the research results reported within this paper \cite{10.1145/3311790.3396656}. URL: http://www.tacc.utexas.edu. 
{MA acknowledges early discussions of related work with Peter Adshead, Mudit Jain, Kaloian Lozanov, Helvi Witek and Hong-Yi Zhang. We also acknowledge many fruitful conversations with Emanuele Berti, Robin Croft, Liina M. Chung-Jukko, Tamara Evstafyeva, Eugene Lim and Ulrich Sperhake.}
Z.W and T.H. are supported by NSF Grants No. AST-2006538, PHY-2207502, PHY-090003 and PHY20043, and NASA Grants No. 19-ATP19-0051, 20-LPS20-0011 and 21-ATP21-0010. MA is supported by a DOE grant DE-SC0021619. This research project was conducted using computational resources at the Maryland Advanced Research Computing Center (MARCC). 

%%%%%%%%%%%%%%%%%%%%%%%%%%%%%%%%%%%%%%%%%%%%%%%%%%%%%%%%%

\appendix

\section{Convergence tests}

 On the initial spatial slice, the Hamiltonian constraint \eqref{eqn:Hamcon}, the momentum constraint \eqref{eqn:Momcon}, and the Proca constraint \eqref{eqn:procaconstraint3d} must be satisfied. We use the conformal-transverse-traceless (CTT) formalism to reduce \eqref{eqn:Hamcon} and \eqref{eqn:Momcon} into elliptical equations of the conformal factor $\psi$ and the extrinsic curvature $K_{ij}$ (see \cite{Baumgarte:2010ndz} chap.~3 and appendix B).

 \begin{figure}[t!]
  \includegraphics[width=0.47\textwidth]{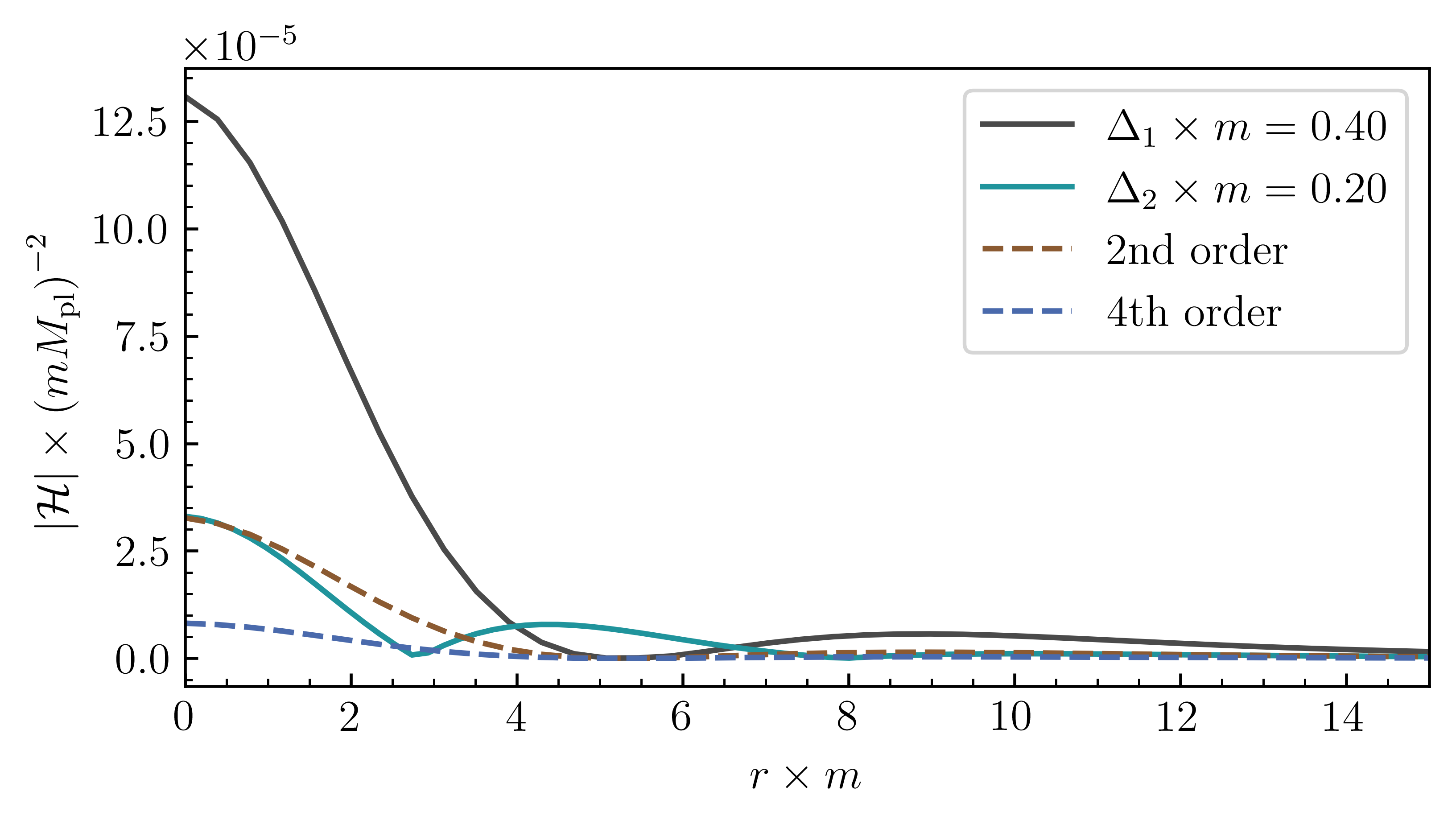}
  \caption{Hamiltonian constraint violation for the densest ($\mathcal{C} = 0.08$) circularly polarized Proca star with two different resolutions $\Delta_1 = 0.40m^{-1}$ and $\Delta_2 = 0.20m^{-1}$. Here $r$ is the radial distance (in code units) from the center of the star. The dashed lines show the predicted Hamiltonian constraint of the high-resolution run for second-order and fourth-order convergence.}
  \label{fig:conv_IC}
\end{figure}

\begin{figure}[ht!]
  \includegraphics[width=0.47\textwidth]{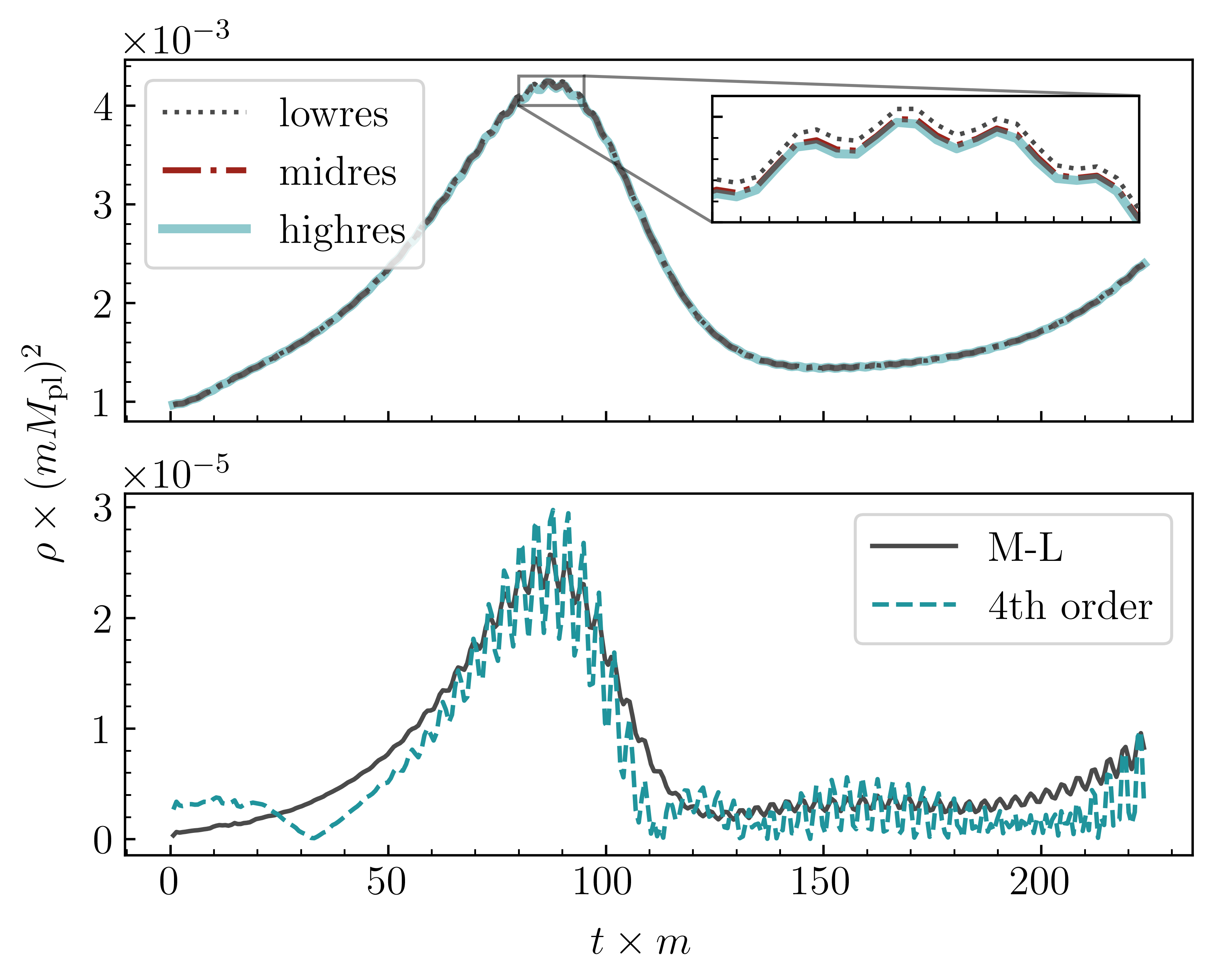}
  \caption{Convergence test for the densest circularly polarized Proca star with three different resolutions. The top panel shows $\rho$ at the center of the star (averaged over a sphere with radius $0.2 m^{-1}$). The bottom panel shows the difference between simulations of medium and low (M-L) resolution (black). The dashed line shows the predicted difference, assuming fourth-order convergence, based on the difference between the high-resolution run and the medium-resolution run.
  }
  \label{fig:conv_cir}
\end{figure}

To show the validity of the solutions obtained from the CTT equations, we performed convergence tests on the initial data of the densest circularly polarized Proca star (generated with $\mu = 0.10m$). In Fig.~\ref{fig:conv_IC}, we show the Hamiltonian constraint $\mathcal{H}$ with two different resolutions. We see that $\mathcal{H}$ from the high-resolution run is smaller than that of the low-resolution run, and is consistent with second-order convergence towards zero. The momentum constraint $\mathcal{M}$ and the Proca constraint $\mathcal{C_\epsilon}$ both behave similarly to $\mathcal{H}$, and are consistent with second-order convergence.

Aside from the initial data test, We also performed a convergence test for the Proca field evolution scheme. Using the same circularly polarized Proca star, We perform three runs with resolutions $\Delta_1 = 0.146m^{-1}$, $\Delta_2 = 0.117m^{-1}$, and $\Delta_3 = 0.098m^{-1}$ at the position of the star. We then plot the energy density $\rho$ at the center of the star (averaged over several cells) in Fig.~\ref{fig:conv_cir}. 
On the top panel, the radial oscillations of the star are visible in the form of changes in its central density. Here, the differences between simulations with different resolutions are negligible compared to the physical density variations.

\section{Numerical Details}

We performed a series of runs, as shown in Figure \ref{fig:MvsR}, where we selected profiles with $\mu/m = 0.01, 0.02, 0.04, \text{and }0.06$ for the hedgehog Proca stars. For the generated linearly polarized $\mu/m = 0.01, 0.02, 0.04, 0.06, 0.087, \text{ and } 0.10$ and circularly polarized Proca stars, we used $\mu/m = 0.01, 0.02, 0.04, 0.06, \text{and } 0.10$.

To determine the $R_{95}$ radius, we integrated the energy density $\rho$ over the volume. We extracted the energy density along the $x$ axes and assumed that these profiles are symmetric for each axis. The mass $M(R)$ is given by the integral:

\begin{equation}
M(R) = 4\pi \int_0^R \rho_{(x)}(r) r^2 dr.
\end{equation}

Next, we determined the radius $R_{95\%}$ that contains 95\% of the total mass: $M(R_{95\%}) = 0.95 M(\infty)$. We calculated the mass based on all three directions $x$, $y$ and $z$ and in most cases, the three radii agreed and did not vary significantly, only when approaching highly relativistic regimes before collapse did we observe noticeable deviations from sphericity. Lastly, this measure is dependent on gauge, so it should be interpreted with care.

%%%%%%%%%%%%%%%%%%%%%%%%%%%%%%%%%%%%%%%%%%%%%%%%%%%%%%%%%%%%%%%%%%%%%%%%%%%%%%%%%%%%%%%
\bibliography{mybib}

\end{document}